\documentclass[12pt]{article}
\usepackage{graphicx}

\newcommand\pubnumber{WSU--HEP--XXYY}
\newcommand\pubdate{\today}

\def\isu{Department of Physics and Astronomy\\
Iowa State University, Ames, IA 50011, USA}

\def\Title#1{\begin{center} {\Large #1 } \end{center}}
\def\Author#1{\begin{center}{ \sc #1} \end{center}}
\def\Address#1{\begin{center}{ \it #1} \end{center}}

\newcommand\pubblock{\rightline{\begin{tabular}{l} \pubnumber\\
         \pubdate  \end{tabular}}}
\newenvironment{Abstract}{\begin{quotation}  }{\end{quotation}}
\newenvironment{Presented}{\begin{quotation} \begin{center} 
             PRESENTED AT\end{center}\bigskip 
      \begin{center}\begin{large}}{\end{large}\end{center} \end{quotation}}

\def\beq{\begin{equation}}
\def\eeq#1{\label{#1}\end{equation}}
\def\eeqn{\end{equation}}

\def\beqa{\begin{eqnarray}}
\def\eeqa#1{\label{#1}\end{eqnarray}}
\def\eeqan{\end{eqnarray}}

\let\bar=\overbar

\def\Dslash{\not{\hbox{\kern-4pt $D$}}}
\def\dslash{\not{\hbox{\kern-2pt $\del$}}}

\def\msb{{\bar{\ssstyle M \kern -1pt S}}}

\begin{document}
\begin{titlepage}
\pubblock

\vfill
\Title{Study of Charmonium Production in Asymmetric Nuclear Collisions by the PHENIX Experiment at RHIC}
\vfill
\Author{Alexandre Lebedev for the PHENIX Collaboration}
\Address{\isu}
\vfill
\begin{Abstract}
The measurement of quarkonia production in relativistic heavy ion
collisions provides a powerful tool for studying the properties of
the hot and dense matter created in these collisions.
To be really useful, however, such measurements must cover a wide
range of quarkonia states and colliding species.
The PHENIX experiment at RHIC has successfully measured $J/\psi, \psi',
\chi_c$ and Upsilon production in different colliding sysyems at various energies.
In this talk I will present recent results from the PHENIX collaboration
on charmonium production in d+Au, Cu+Au and U+U collisions at 200 GeV/c.
 
\end{Abstract}
\vfill
\begin{Presented}
The 7th International Workshop on Charm Physics (CHARM 2015)\\
Detroit, MI, 18-22 May, 2015
\end{Presented}
\vfill
\end{titlepage}
\def\thefootnote{\fnsymbol{footnote}}
\setcounter{footnote}{0}
%

\section{Introduction}

Dissociation of quarkonia by color screening in deconfined
matter is predicted to be different for different states.
Loosely bound states will melt first, and successive suppression of
individual states can provide an effective thermometer of the QGP.
However, there are many competing processes in nucleus-nucleus collisions:
cold nuclear matter effects, color screening, initial state effects,
regeneration, feed-down, and so on. 
Thus, in order to have a clear picture of what happens during relatistic nucleus-nucleus
collisions, we need measurements for different energies,
colliding species, and quarkonium states.
In this respect, asymmetric nucleus-nucleus collisions can be very useful 
in understanding the importance of different processes contribuing to quarkonia production.

\section{d+Au collisions}

\begin{figure}[htb]
\centering
\includegraphics[width=5.5cm]{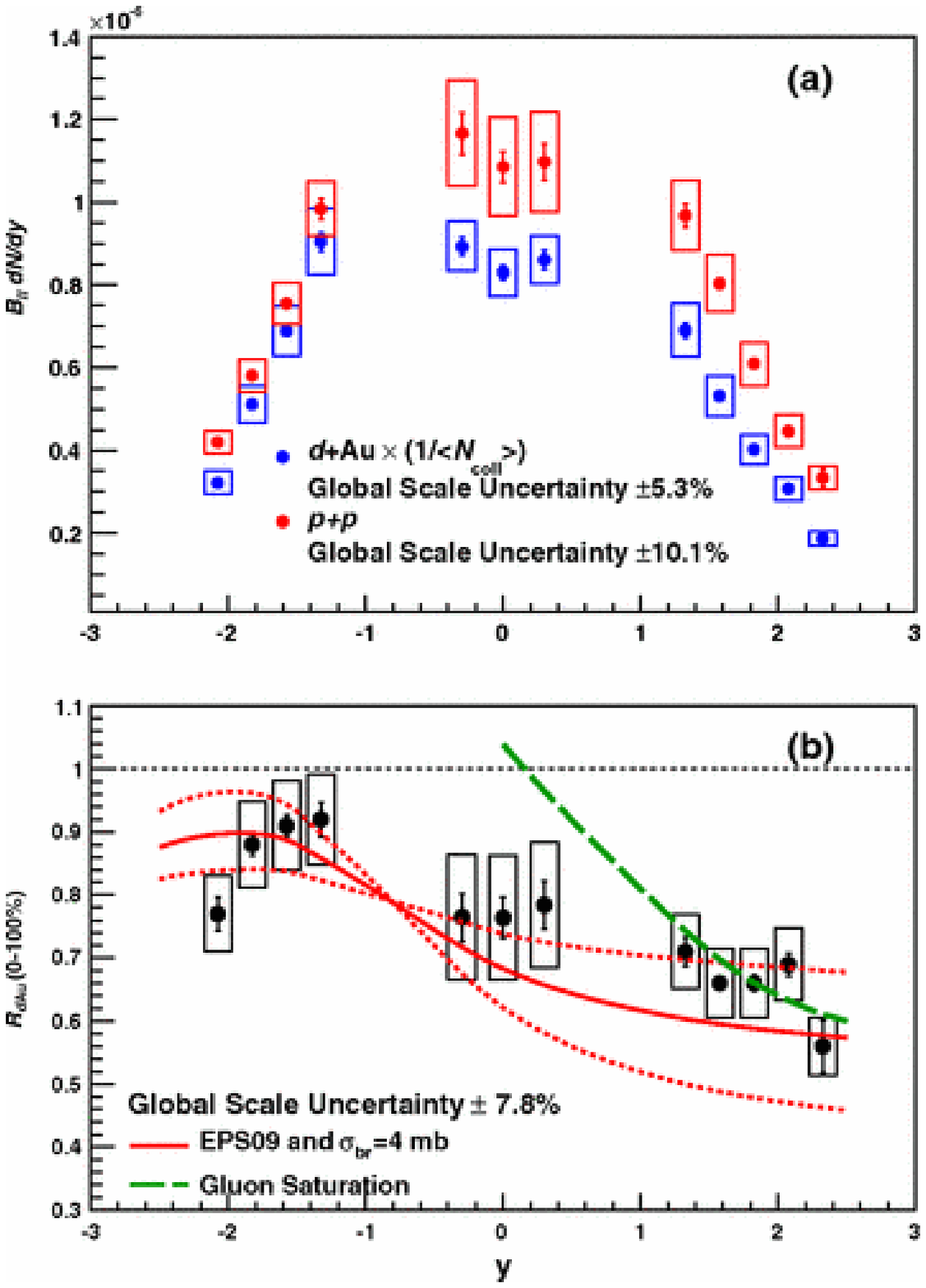}
\includegraphics[width=5.2cm]{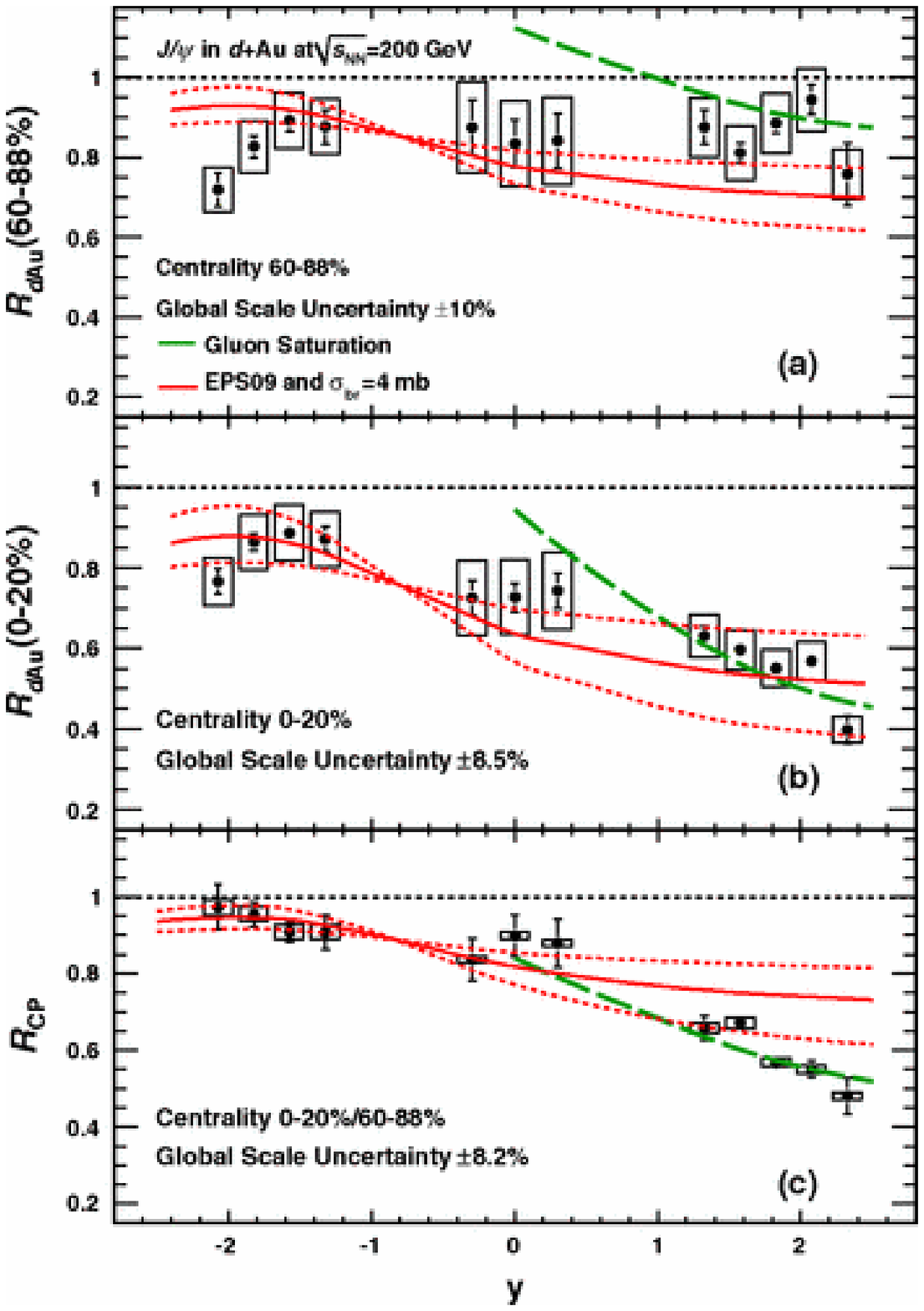}
\caption{Rapidity depencence of $J/\psi$ yield in p+p and d+Au collisions (top left),
and of $R_{dAu}$ (bottom left). 
Rapidity dependence of $R_{dAu}$ for different centralities (right).
}
\label{fig:fig1}
\end{figure}

The PHENIX experiment has measured $J/\psi$ production in p+p and d+Au collisions at 200GeV.
at forward, backward, and central rapidities~\cite{prl107}.
Fig.~\ref{fig:fig1}(top left) shows the $J/\psi$ invariant yields in p+p
and d+Au collisions as a function of rapidity, integrated over centrality
(0\%–100\%). The error bars (boxes) represent point-to-point
uncorrelated (correlated) uncertainties.

The cold nuclear matter effects are quantified by calculating
the nuclear modification factor $R_{dAu}$, which is defined as
the ratio of $J/\psi$ yield in d+Au collisions to $J/\psi$ yield in p+p collisions,
corrected for the number of binary collisions $N_{coll}$. $N_{coll}$ 
is derived using a Glauber calculation  (see \cite{glauber} for details).

As expected, nuclear modification factor $R_{dAu}$ exibits rapidity asymmetry
in d+Au collisions, as is shown in Fig.~\ref{fig:fig1}(bottom left).
Forward (deuteron going) rapidity shows more suppression than central
and backward (Au going) rapidity.

This difference can have many possible explanations, including nuclear 
breakup and gluon shadowing.
A model~\cite{Ramona} which uses EPS09 nPDF and breakup cross-secton $\sigma_{br} = 4 mb$ 
shows reasonable agreement with the data. Predictions of this model are shown in Fig.~\ref{fig:fig1}
by red curves. 
A second class of calculations incorporates gluon saturation
effects at small-x~\cite{Kharzeev}, and is shown in Fig.~\ref{fig:fig1} by green lines.
This model the data well at forward rapidity, but fails to reproduce them at backward rapidity.

However, if one looks at centrality dependence of $R_{dAu}$ the picture
becomes less unambiguous. 
Fig.~\ref{fig:fig1}(right) shows rapidity dependence of $R_{dAu}$ for peripheral (top) and
central (middle) d+Au collisions, as well as their ratio $R_{CP}$ (bottom).
For peripheral collisions, the $R_{dAu}$ ratio shows a mild
suppression, roughly independent of rapidity, within the
systematic uncertainties of approximately 15\%. 
For central collisions $R_{dAu}$ indicates a much larger suppression for
$J/\psi$ at forward rapidity.
As one can see, the model which uses EPS09 nPDF and breakup cross-secton $\sigma_{br} = 4 mb$~\cite{Ramona},
fails to describe the $R_{CP}$ measurement at
forward rapidity. No parameter choice of the
EPS09 nPDF set and of $\sigma_{br}$ is able to describe the rapidity
and centrality dependence of the data. Gluon saturation model~\cite{Kharzeev} describes
centrality dependence well at forward rapidity, but fails at other rapidities.

\begin{figure}[htb]
\includegraphics[width=15cm]{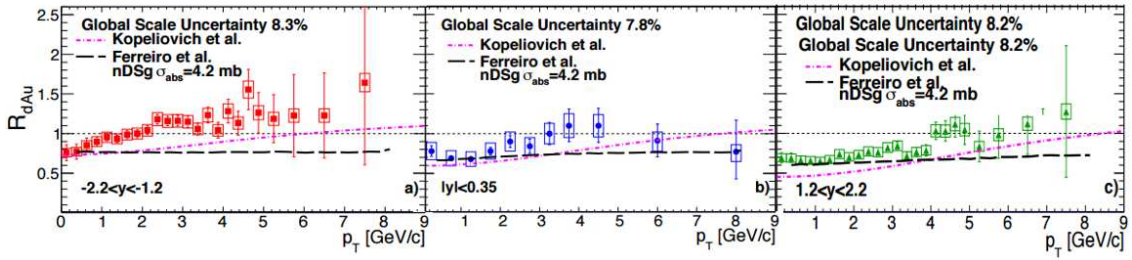}
\caption{Transverse momentum dependence of $R_{dAu}$ for three different rapidity ranges.}
\label{fig:fig2}
\end{figure}

Fig.~\ref{fig:fig2} shows transverse momentum dependence of $R_{dAu}$ for different
rapidities.
As one can see, at all rapidities, $R_{dAu}$ rises up to $5GeV/c$. 
A model which includes shadowing + $\sigma_{br}$, shown by red dashed curves, does not match the trend.
The model by Kopeliovoich et al.~\cite{Kopeliovich}, which includes Cronin effect and $\sigma_{br}$
qualitatively matches the shape of observed dependence.
Largest disagreement with theories is observed at backward rapidity.

\begin{figure}[htb]
\centering
\includegraphics[width=10cm]{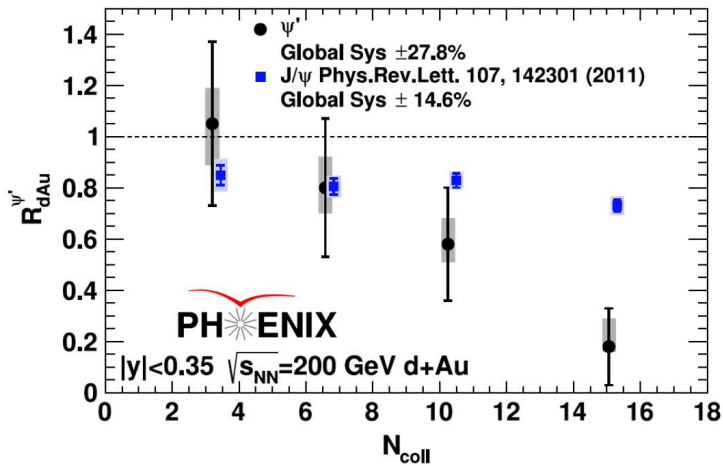}
\caption{$R_{dAu}$ for $\psi$' and $J/\psi$ as a function of centrality.
Note that the $J/\psi$ $R_{dAu}$ plotted here is not corrected for
$\chi_c$ and $\psi$' feed-down, and the Ncoll values are shifted slightly to
aid in clarity.
}
\label{fig:fig3}
\end{figure}

Fig.~\ref{fig:fig3} shows $R_{dAu}$ for $\psi$'~\cite{prl111}.
Unexpectedly, $\psi$' is ~3 times more suppressed in most 
central collisions than $J/\psi$, and has very different trend with $N_{coll}$ compared to $J/\psi$.
Reference~\cite{Arleo} presents a model that explains the lower
energy E866/NuSea and NA50 relative $\psi$'/$J/\psi$ suppression results using an expanding
color neutral $c\bar{c}$ pair. As the $c\bar{c}$ expands, it has an increased
nuclear absorption owing to its larger physical size. Once
the time spent by the $c\bar{c}$ pair traversing the nucleus ($\tau$)
becomes larger than the $J\psi$ formation time, the $\psi$' will
see a larger nuclear absorption owing to its larger size.
This idea is tested at RHIC energies by calculating the
average proper time spent in the nucleus by the quarkonia
(or $c\bar{c}$ precursor).

\begin{figure}[htb]
\centering
\includegraphics[width=10cm]{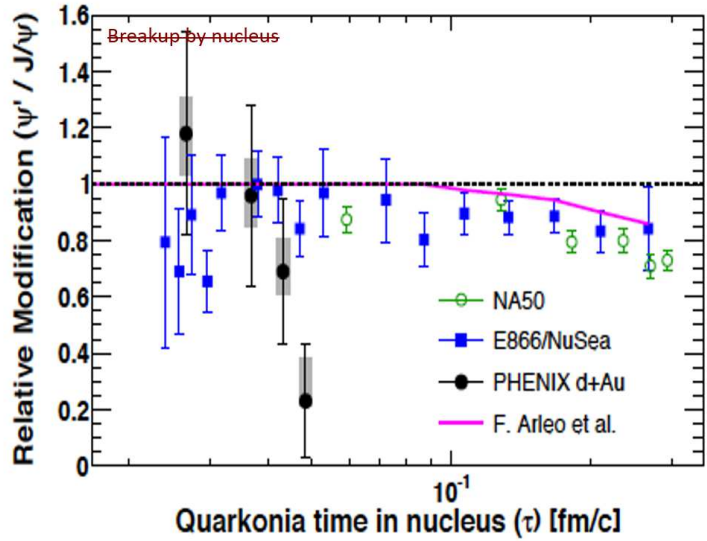}
\caption{Relative modification $\psi$'/ $J/\psi$ as a function of quarkonia time in nucleus.
The curve is a calculation by Arleo et al.~\cite{Arleo} discussed in the text.
}
\label{fig:fig5}
\end{figure}

Fig.~\ref{fig:fig5} shows
nuclear crossing time in d+Au for different collision energies.
Universal trend with $dN_{ch}/d\eta$ for several 
systems, up to 200 GeV is observed.
The solid curve in Fig.~\ref{fig:fig5} is the calculation by Arleo et al.~\cite{Arleo}, which is
consistent with the trends observed by E866/NuSea and NA50.
However, the PHENIX data show very different $\tau$ dependence.

The values of $\tau$ for the PHENIX data are similar to
the  $c\bar{c}$ formation and color neutralization time of $\sim0.05 fm/c$, and well below the $J\psi$ formation time of 
$\sim0.15 fm/c$~\cite{Arleo}. 
Therefore the model cannot explain the
strong differential suppression of the $\psi$' in the PHENIX
data.

\section{Cu+Au collisions}

\begin{figure}[htb]
\centering
\includegraphics[width=10cm]{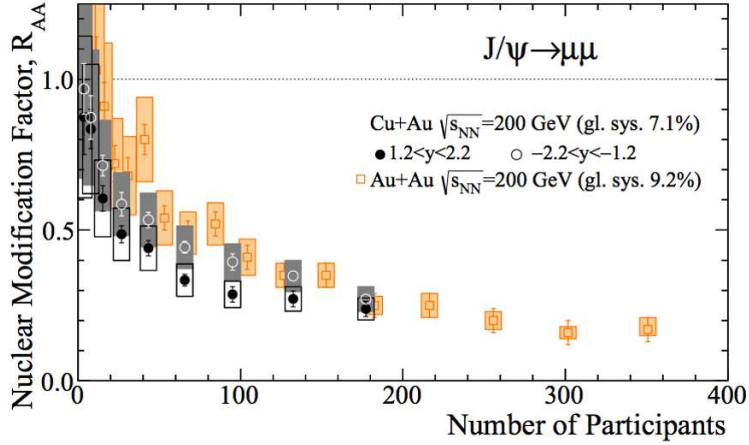}
\caption{$R_{AA}$ vs. centrality for Cu+Au and Au+Au collisions.}
\label{fig:fig1cuau}
\end{figure}

Fig.~\ref{fig:fig1cuau} shows $J/\psi$ $R_{AA}$ vs. centrality for Cu+Au collisions~\cite{cuau}
as open black circles (backward rapidity) and solid black circles (forward rapidity).
$R_{AA}$ for Au+Au collisions is shown on the same plot for comparison as orange points.
Observed suppression is somewhat smaller in peripheral Cu+Au collisions, compared to
Au+Au collisiosn, but becomes the same with increasing centrality.
Higher suppression is observed in the region of lower particle 
density (forward rapidity), similar to d+Au collisions.
Debye screening would go in the other direction.

\begin{figure}[htbp]
\centering
\includegraphics[width=10cm]{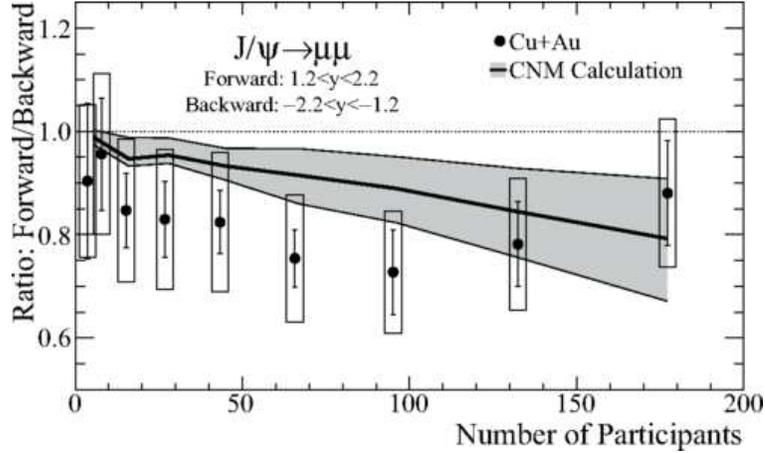}
\caption{$R_{AA}$ ratio for Cu-going/Au-going directions in Cu+Au collisions.
}
\label{fig:fig2cuau}
\end{figure}

\begin{figure}[htbp]
\centering
\includegraphics[width=10cm]{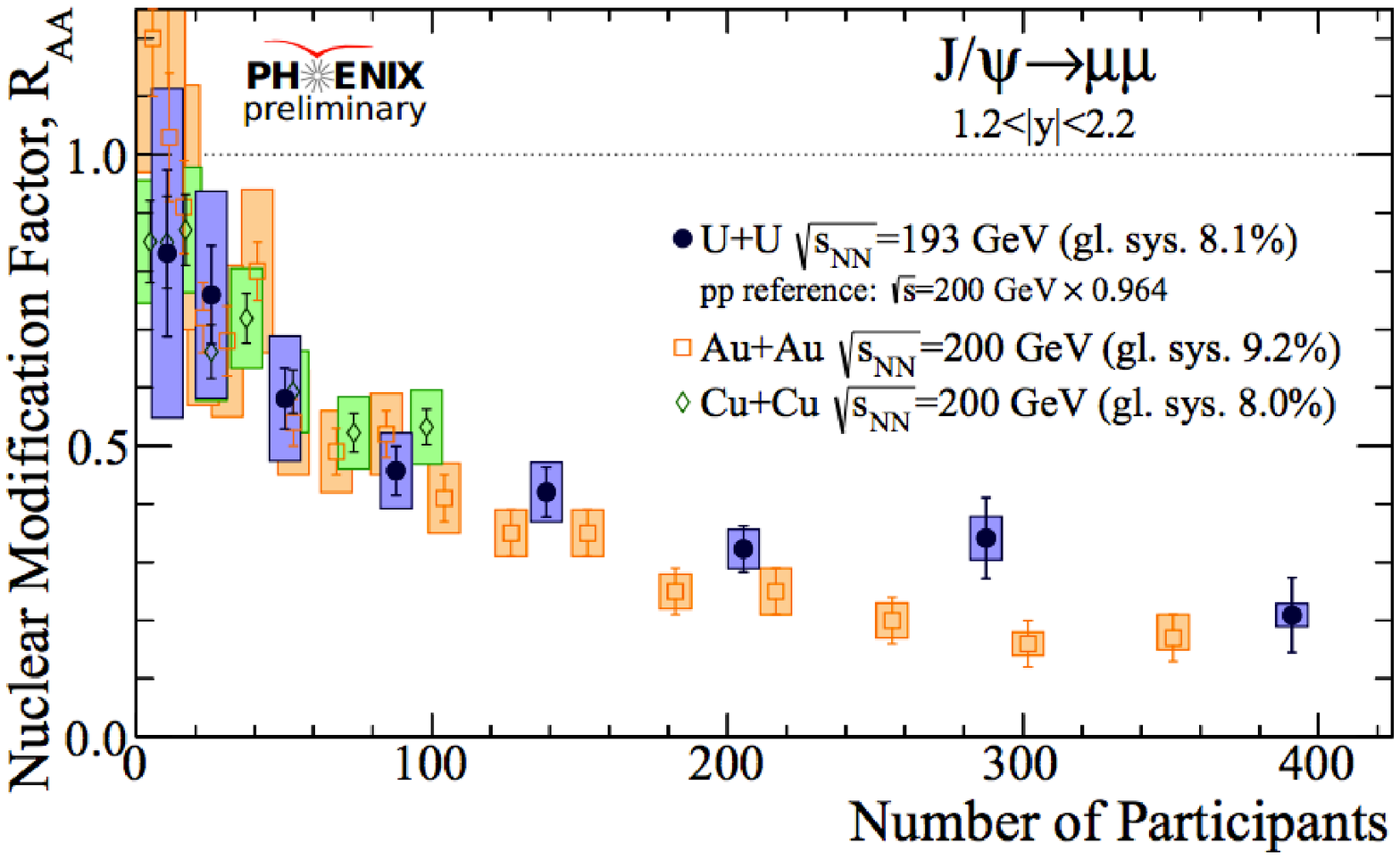}
\caption{$R_{AA}$ vs. centrality for U+U collisions (black circles) compared to
Au+Au and Cu+Cu collisions (open symbols).}
\label{fig:fig1uu}
\end{figure}

Fig.~\ref{fig:fig2cuau} shows a ratio of $R_{AA}$ for Cu-going and Au-going directions.
This ratio
has the advantage of reduced systematic uncertainties.
Observed ratio decreases with centrality. 
The 20\% – 30\%
difference in suppression between forward and backward
rapidity $R_{AA}$ evident in Fig.~\ref{fig:fig2cuau} could be due to hot matter
effects, CNM effects, or a combination of both.
Also shown in this figure as solid line is a model~\cite{ref43} which estimates
the contribution from cold nuclear
matter (shadowing). The grey band in this figure represents the extreme nPDF parameter sets for the model.

As can be seen form Fig.~\ref{fig:fig2cuau} 
The difference between forward (Cu-going) and backward
(Au-going) $J/\psi$ modification is found to be comparable
in magnitude and of the same sign as the expected
difference from shadowing effects.

\section{U+U collisions}

Collisions of deformed uranium nuclei produce a wide variation in energy density within 
the same colliding system.
MC studies show~\cite{prc76} a possibility of selecting experimentally 
tip-tip collisions by selecting high multiplicity, but low flow events.  
In tip-tip collisions $T/T_C$ could reach above 2 \cite{prc84}, at which tempertaure $\Upsilon(1S)$ 
could dissociate.

The PHENIX collaboration recently measured $J/\psi$ production in U+U collisions at 200GeV.
Fig.~\ref{fig:fig1uu} shows preliminary results of $J/\psi$ $R_{AA}$ as a function
of centrality for U+U collisions (black circles) compared to
Au+Au and Cu+Cu collisions.
Qualitatively similar $J/\psi$ suppression is observed from Cu+Cu to U+U collisions.
Somewhat weaker suppression in most central U+U collisions may indicale higher role of
coalescence processes in uranium-uranium collisions.

\section{Conclsions}

In d+Au collisions $J/\psi$ nuclear modification factors at forward rapidity as a
function of centrality cannot be reconciled with a picture of
cold nuclear matter effects (nPDFs and $\sigma_{br}$) when an
exponential or linear dependence on the nuclear thickness
is employed.

In Cu+Au collision, the Cu going side is more suppressed than Au
going side, consistent with CNM effects (shadowing).

The magnitude and trend of $\psi(2s)$ suppression in nuclear collisions
is quite different from ithat of $J/\psi$. Nuclear crossing time does not explain
the data.

$J/\psi$ $R_{AA}$ is qualitatively consistent between 
different colliding systems, from Cu+Cu to U+U. 
$\sim$25\% differences could be due to expected variations in the CNM and QGP effects.



\end{document}